%% file: head.tex
\documentstyle[psfig]{mn}
\newcommand{\bgr}{\bibitem[\protect\citename{dummy }1893]{dum}}
\newcommand{\etalc}{et~al.}
\begin{document}
\title{High spatial resolution observations of CUDSS14A:  a Scuba-selected 
Ultraluminous galaxy at high redshift}

\author[W.K. Gear {\it et al.}]
{W. K. Gear$^1$,  
S.J. Lilly$^{2,\star}$, 
J.A. Stevens$^3$,
D.L. Clements$^1$,
T.M. Webb$^2$, \cr
S.A. Eales$^1$ \& 
L. Dunne$^1$
\\
$^1$ Department of Physics and Astronomy, Cardiff University, PO Box 913, 
Cardiff CF2 3YB\\
$^2$ Department of Astronomy, University of Toronto, 60 St. George 
Street, Toronto, Ontario M5S 3H8, Canada\\
$^3$ Mullard Space Science Laboratory, University College London, 
Holmbury St. Mary, Dorking, Surrey RH5 6NT \\} \date{draft 1.0}

\maketitle
\begin{abstract}
We present a high-resolution millimetre interferometric image of the brightest
SCUBA-selected galaxy from the Canada-UK deep SCUBA survey (CUDSS). We
make a very clear detection at 1.3 mm, but fail to
resolve any structure in the source. The interferometric position is
within 1.5 arcsec of the SCUBA 850 $\mu$m centroid, and also within
1.5 arcsec of   
a 44 $\mu$Jy radio source and a very faint, extremely red galaxy which we 
had previously identified as the submillimetre source. We also present 
new  optical and infrared imaging, and infrared spectroscopy of this 
source. We model the overall spectral energy distribution and conclude 
that it lies within the redshift range 2$< z < 4.5$. The submm/FIR 
luminosity of CUDSS14A is very weakly dependent on 
redshift within the constrained range, and is roughly 
4$\times$10$^{12}$ L$_\odot$ (for H$_0$=75 and an assumed Arp220-like spectrum), which implies a 
star-formation rate $\sim$1000 M$_\odot$ yr$^{-1}$. We derive an
approximate gas mass of $\sim 10^{10}$ M$_\odot$ which would imply the
current star-forming activity cannot be sustained for longer than about 10
million years. With the present data however we are unable to rule out a 
significant AGN contribution to the total luminosity.

\end{abstract}

\begin{keywords}
Galaxies, SCUBA sources, galaxies, high-redshift
\end{keywords}
\footnotetext{E-mail: gear@astro.cf.ac.uk\\
$\star$Visiting Observer at the Canada-France-Hawaii Telescope 
 operated by the NRC of Canada, the CNRS of France and the University of 
Hawaii}
\input{body}

\section*{ACKNOWLEDGMENTS}

JAS and DLC acknowledge the support of  PPARC PDRAs. We thank
R.Moreno and A.Dutrey for help with the IRAM data reduction. 
SJL's research is supported by the NSERC of Canada
and by the Canadian Institute for Advanced Research. We thank the referee for helpful comments.

\bsp
\end{document}

%% file: body.tex
\section{Introduction}

The commissioning of SCUBA on the JCMT has revolutionized the field of
observational cosmology and in particular the study of galaxy evolution,
arguably making as big an impact as the release of the Hubble Deep Field
did a few years earlier. 

Something of the order of 50 galaxies have now
been discovered in the various deep SCUBA surveys (Smail et al 1997,
Hughes et
al 1998 , Barger et al 1998, Eales et al 1999) and at least 30 percent of
the
far-infrared/submillimetre background discovered by COBE (Puget et al
1996,
Fixsen et al 1998) has been resolved. Considerable effort is
now
being dedicated at ground-based telescopes in the radio, infrared and
optical, as well as Satellite observatories such as the Hubble Space Telescope,XMM-Newton and Chandra, to following up these
sources, making identifications and obtaining redshifts, so that the
history of star-formation in the Universe can be properly characterised
free of the obscuring effect of dust.

Accurate positions are of course essential for identifications and 
follow-up observations, and the 15 arcsec SCUBA beam at 850 $\mu$m allows an uncertainty of several arcseconds even on a high signal-to-noise detection.
Deep VLA imaging has been quite successful at providing identifications and accurate positions of a number of SCUBA-selected sources (Smail et al 2000, Richards 1999). However,  $\leq 50 \%$ of SCUBA sources are detected with the VLA, even at $\sim$ 10$\mu$Jy sensitivity levels, and in some cases even with a VLA detection there may be some ambiguity in the identification (Richards 1999, Downes et al 1999). The only way to truly {\it confirm} the identification is to make a higher resolution millimetre or submillimetre image using an interferometric array (Frayer et al 1998,1999,2000; Downes et al 1999).
 
In this paper we present the results of an observation with the IRAM
interferometer on Plateau du Bure of CUDSS14A, the brightest source
discovered in the CUDSS as reported by Eales et al (1999), along with new   
high resolution J and K images obtained with the CFHT Adaptive Optics system
PUEO which we combine with archive HST images at 606 and 814 nm (roughly V and I), and also new infrared spectroscopy obtained 
on the United Kingdom Infrared Telescope (UKIRT).

\section{Observations}

\subsection{IRAM Observations} 

The IRAM data were taken on 25, 26 and 28
November 1998. The array was in the 5D configuration, and the source was
tracked for 8.4, 3.3 and 10.6 hours on each night respectively. Flux
calibration was performed against CRL618, 3C273 and MWC349 with 1637+574
and 1418+546 being used as phase calibrators. The observations were
centred on the nominal SCUBA position of (J2000) 14 17 40.3 +52 29 08 and
observations were made at 1.3 mm (238.5 GHz) and 2.8 mm (105 GHz) 
simultaneously. The data were correlated and reduced at the IRAM
headquarters in Grenoble using the standard software.

The 1.3 mm map is shown in Figure 1 (we note that Bertoldi et al 2000 have produced an essentially identical map from the same data). The object
is very clearly detected at a level of 6 $\sigma$, and the positional
centroid is 14 17 40.37 +52 29 06.8, or 0.6 arcsec in RA and 1.2
arcsec in declination away from the nominal SCUBA position. We estimate 
the total uncertainty of the IRAM position to be $\pm$ 0.3 arcsec (see 
Downes et al
1999 for a detailed discussion of the different factors affecting the
positional uncertainty of IRAM interferometer maps). There is no evidence for any extension beyond the beam. We also examined the data for 
any evidence of spectral line emission but found none, however this does not place any meaningful limits on the redshift.

There was no
detection  at 2.8 mm. The 1.3 mm flux and 2.8 mm upper limit are shown in Table 1, along with the SCUBA and ISO measurements. The overall flux distribution, which is very similar to other SCUBA-selected galaxies (e.g. Ivison et al 1999), is plotted in Figure 2.

\begin{figure}

\setlength{\unitlength}{1in}
\begin{picture}(2.5,2.5)
   \includegraphics{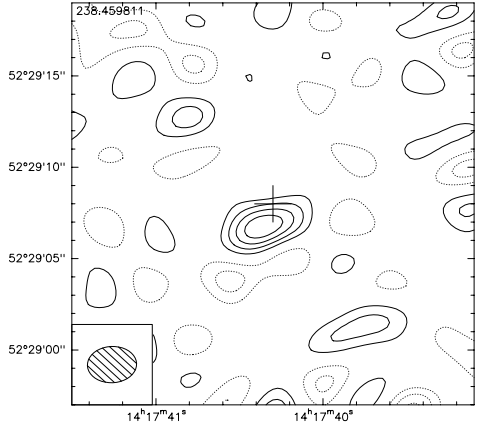}
\end{picture}
   \caption{The 1.3 mm map of CUDSS14A, contours are in steps of 0.5 mJy,
the 1 sigma uncertainty, and extend from -2 $\sigma$to  +5$\sigma$.The
cross marks the nominal SCUBA position on which the observation was 
centred. The effective beam shown is 2.7 $\times$ 2.0 arcseconds with the major axis at a PA of 93 degrees.  } 
\end{figure} 

\begin{table}
\small
\centering                                                                     
\def\baselinestretch{1}                                     
\caption[dum]{\small Radio, Submm and ISO photometry}
\begin{tabular}{lcl}\hline
Wavelength & Flux (mJy) & reference\\ \hline
6 cm & 0.044 $\pm$0.004 & Fomalont \etalc 1991 \\
2.8 mm & 3$\sigma <$0.54& this paper \\
1.3 mm & 2.94 $\pm$0.49 & this paper\\
850 $\mu$m & 8.8 $\pm$1.1 &Eales et al 1999  \\
450 $\mu$m & 3$\sigma <$31 &Eales et al 1999 \\
15 $\mu$m & 3$\sigma <$ 0.2 & Flores et al 1998a \\
7 $\mu$m &3$\sigma <$ 0.15 &  Flores et al 1998b \\
\hline
\end{tabular}
\end{table}

\begin{figure}

\setlength{\unitlength}{1in}
\begin{picture}(2.9,2.8)
   \includegraphics{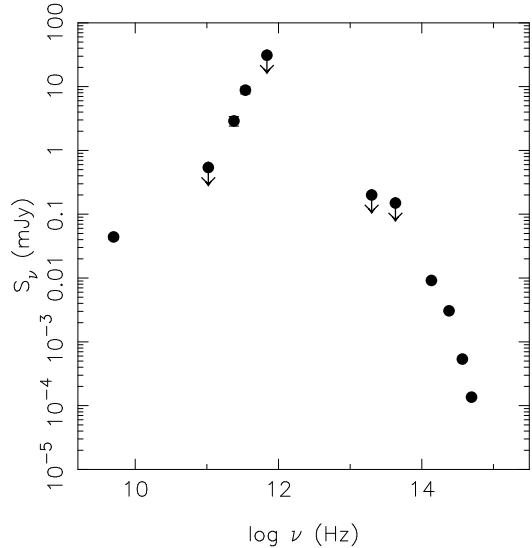}
\end{picture}
   \caption{The overall radio through optical observed flux
distribution of CUDSS14A.} 
\end{figure}

\begin{table}
\small
\centering                                                                     
\def\baselinestretch{1}                                     
\caption[dum]{\small Multi-aperture optical/infrared photometry (AB mags)}
\begin{tabular}{lccc}\hline
Filter & 1 arcsec & 2 arcsec & 4 arcsec\\ \hline
F606W(HST) & 27.132$\pm$0.314 & 27.011$\pm$0.351 & 26.063$\pm$0.263 \\
F814W(HST) & 25.228$\pm$0.127 & 25.023$\pm$0.128 & 24.574$\pm$0.128 \\
J(CFHT)    & 23.230$\pm$0.094 & 22.866$\pm$0.124 & 22.686$\pm$0.215 \\
K'(CFHT)    & 21.423$\pm$0.047 & 21.195$\pm$0.072 & 21.495$\pm$0.197 \\
\hline
\end{tabular}
\end{table}

\subsection{New CFHT Observations}

Observations were made with the PUEO Adative Optics system on 
CFHT during the nights of April 27-28 1999, using an R=15.3 guide 
star 26 arcsec WNW of the target. The detector was a 1024$^2$ HgCdTe 
array sampling the image at 0.035 arcsec/pixel. Secure detections were obtained in a total of 150 minutes integration in J and 100 in K'.

\begin{figure}
\setlength{\unitlength}{1in}
\begin{picture}(3.0,3.0)
   \includegraphics{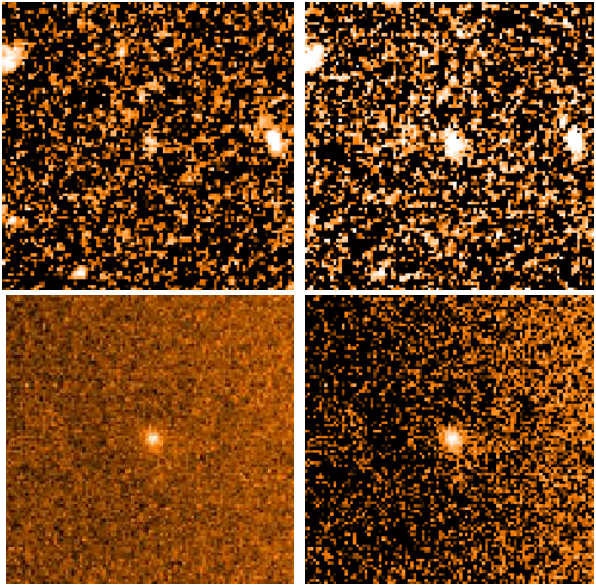}
\end{picture}
   \caption{Images at F606W, F814W, J and K' clockwise from top left respectively of 
CUDSS14A. The images are all co-centred on the extremely red optical 
source, and are all 10 $\times$ 10 arcseconds.} 
\end{figure} 

The new infrared images have been co-registered with archival HST images 
in F606W and F814W (the source is in the so-called 'Westphal' survey 
field), resampling to a pixel scale of 0.1 arcsec/pixel. A montage of the 
F606W, F814W, J and K' images is shown in Figure 3. The morphology of the source appears compact and indistinct. The FWHM of the images are about 0.4 arcsec in all of the 
bands, substantially larger than the PSF in the HST data. The PSF is 
poorly constrained in the Adaptive Optics images but is most likely 
substantially smaller than 0.4 arcsec. 

The brightness of CUDSS-14A in F606W, F814W J and K' in 
apertures of 1, 2 and 4 arcsec diameter are listed in Table 2. 
The colours are broadly consistent in the different apertures but 
with a trend to be bluer in larger apertures. This new photometry 
agrees well with earlier measurements in larger apertures by Lilly et al 
(1999). Optical spectroscopy was also attempted but no detection was made.

\subsection{UKIRT Observations}

CUDSS14A was observed with the CGS4 spectrometer on the nights of April
13 and 16 1999 using the 300 mm focal length camera and 40 lines/mm
grating. The 2.4 arcsec slit was centred on the radio position, and the
observing and data reduction procedures described by Eales and Rawlings
(1993) were used. In
the K and H bands we observed in first order ($\lambda/\delta \lambda = 
400$) 
for 160 minutes, and in J we observed in second order ($\lambda/\delta 
\lambda = 800 $) for 80 minutes.

Continuum was detected at K,  but not at J or H. No spectral lines were seen
in any of the bands and we estimate the upper limits on the lines as $\rm 1.1
\times
10^{-19}\ W\ m^{-2}$ in the region 2.02-2.3 $\mu$m, $\rm 1.3 \times
10^{-19}\ W\ m^{-2}$ in the region 1.5-1.8 $\mu$m, and $\rm 2.8 \times
10^{-19}\ W\ m^{-2}$ in the region 1.1-1.3 $\mu$m. To derive these limits
we have assumed that a spectral line would have a spectral width
equivalent to a velocity dispersion of 200 km $s^{-1}$. These limits are
slightly more conservative than 3$\sigma$ limits, since we have enough
spectral range that we might expect a 3$\sigma$ fluctuation by chance. The
limits have been chosen to be slightly larger than the largest feature in
each spectral range.

\section{DISCUSSION}

\subsection{The identification}

The IRAM position is 1.4 arcsec from the nominal SCUBA position and is also
only  1.2 arcsec from the 5 GHz position of Fomalont et al (1991) and 1.5 
arcsec from the optical ID (Lilly et al 1999).  Since there is no other 
candidate in each  band than the one we show and
all the positions lie easily within their respective 2$\sigma$ astrometric 
uncertainties of each other (in fact the
IRAM, radio and optical positions are only just outside their respective
1-$\sigma$ uncertainties) we believe the identification to be extremely robust.

\subsection{The Redshift}

In Lilly et al (1999) we suggested a redshift
in the range z=2-3, based on the rather uncertain optical and infrared
colours available at that time. We can now use our new data to provide further photometric constraints. 

\subsubsection{Radio and Submillimetre}

Carilli and Yun (1999; hereafter CY99) suggested that the strong radio-FIR
correlation could be used to provide constraints on the redshifts of SCUBA
survey sources. The original CY99 relationship is based on a 
model, but in a follow-up paper (Carilli and Yun 2000, hereafter CY00) they 
used observational data from 17 galaxies observed in the submillimetre by 
Lisenfield, Isaak and Hills (1999) to produce an empirical relationship, 
which tends to result in somewhat lower redshifts than their original one. 
Both these results have now been superseded however by  Dunne 
et al  (2000) who have made a SCUBA survey of 104 nearby galaxies. Using Dunne
et al's  empirical relationship between 850 $\mu$m and 1.4 GHz flux density,
 including the intrinsic scatter to derive an uncertainty, the
measured 1.3 mm to 850 $\mu$m spectral index of 3.2 and the 1.4 to 5 GHz 
radio spectral index of -0.2$\pm$0.3, we obtain z$\sim$2.2$\pm$0.5 for CUDSS14A. For completeness we note that CY00 would result in z$\sim$2.9$\pm$0.7 and 
CY99 in z$\sim$5.0$\pm$1.0. The rough consistency of the CY00 and Dunne et al 
results and the inconsistency of the CY99 prediction demonstrates both the 
uncertainty in this technique (see also Blain 1999) but also the manner in 
which it is rapidly improving.

We can also use the fact that we have upper limits at 450 $\mu$m and 2.8
mm to provide  further constraints. To do this we take
two extreme template spectra, namely the nearby starburst galaxy M82,
which has a dust temperature of 48K and is optically thin throughout the
FIR and submm region (Hughes et al 1994), and Arp220, which has a fitted
dust temperature of 62K and is optically thick shortward of 200 $\mu$m (see
 e.g. Downes, Solomon \& Radford 1993). 

As demonstrated in Figure 4, CUDSS14A {\it cannot} be at low redshift, for
either template, otherwise we would have detected it at 450 $\mu$m.
Conversely it is unlikely to be at very high redshift, for either
template, otherwise
we would have detected it at 2.8 mm. In fact the detections and upper
limits constrain the redshift to the range  2$< z < 5$.

\begin{figure*}

\setlength{\unitlength}{1in}
\begin{picture}(3.0,3.0)
   \includegraphics{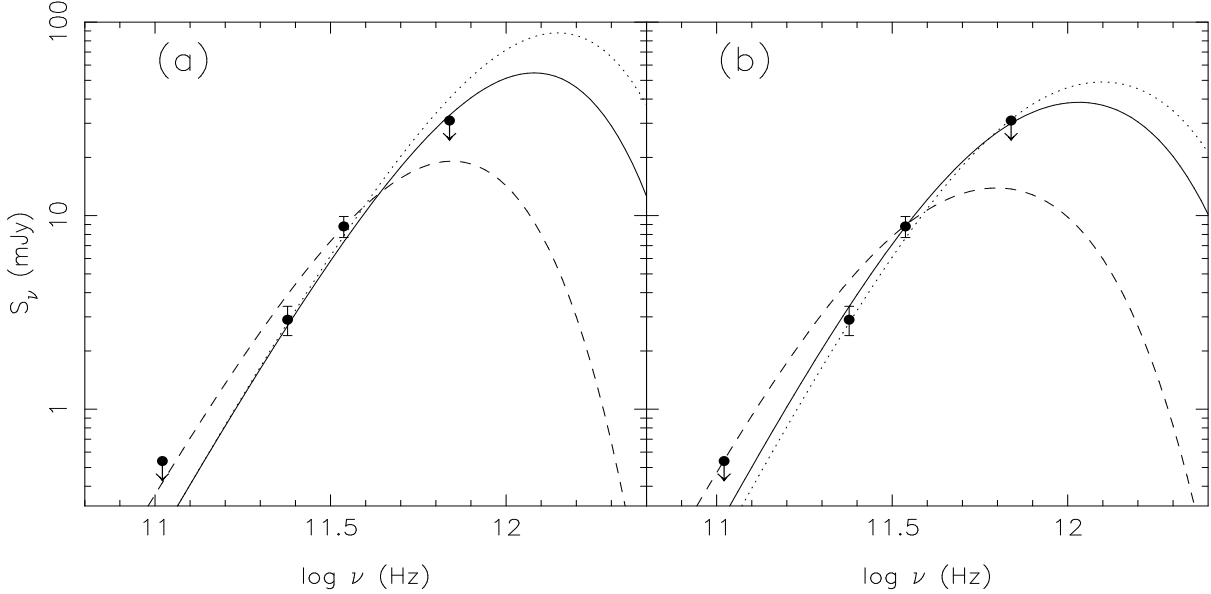}
\end{picture}
   \caption{Comparison of the mm/submm spectrum of CUDSS14A with that of
(a) M82 and (b) Arp220 at z=2.0 (dots), 2.5 (solid) and 5 (dashed) with
arbitrary shifts in absolute flux level. Note that the M82 spectrum is not
consistent with the data for z=2.0 but is just consistent for z=5, whereas
Arp220 is just consistent at z=2.0 but is not consistent for z=5. Thus for
these templates we can constrain the redshift to the range 2.0 to 5.0.  } 
\end{figure*} 

It would be possible to allow a lower redshift if the emitting
dust were much colder. In order to be consistent with the 1.3 mm detection and 450 $\mu$m upper limit however, for z$\leq$1 the temperature would have to be $\leq$ 20K. Although this would reduce the source luminosity somewhat (depending on the unobserved FIR spectral shape), it also increases the required gas and dust mass to $\geq 10^{11}M_\odot$, which is rather excessive. Interestingly, the 450/1300 $\mu$m flux constraint for CUDSS14A is identical to that found for HDF850.1 by Downes et al (1999) and they reach the same conclusion, that the source is unlikely to be at such low redshift. We also note that no sources as cold as 20K are known at low redshift and furthermore, that higher luminosity sources also tend to be warmer. Even at the lower end of any possible luminosity range it seems very unlikely to us that such a luminous object would be very cold. 

\subsubsection{Optical and Infrared}

CUDSS14A is an extremely red object (ERO) with $(V-K)_{AB} \sim 6$ 
corresponding to a Vega-normalized colour of $(V-K) \sim 8$. Other EROs 
have been identified with SCUBA sources (Smail et al 1999), and Barger, 
Cowie and Richards (2000) have shown that many very red radio 
source identifications are detectable with SCUBA. An interesting 
consequence of the very red colours of a source like CUDSS14A is that 
such galaxies would rapidly become extremely faint if placed at higher 
redshifts (Dey et al 1999). This is illustrated in Figure 5 where we show the 
expected V, I, J and K magnitudes as a function of redshift assuming that the
real source is (a) at z=2 and (b) at z=4. In the former case, the source would 
become very hard to detect 
at wavelengths below 1$\mu$m at redshifts as low as 3. The implications 
of this are that SCUBA sources for which no optical identification can be 
found even to extremely faint levels such as reached by the Hubble Deep 
Field are not neccessarily at very high redshifts, z$\gg$3.

\begin{figure}

\setlength{\unitlength}{1in}
\begin{picture}(2.4,2.4)
   \includegraphics{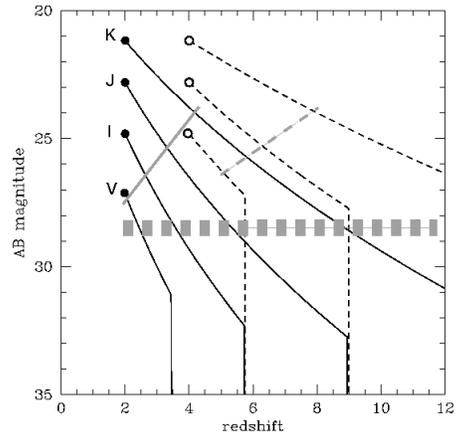}
\end{picture}
   \caption{The expected variation of the V, I, J and K magnitudes 
(in AB) with 
redshift of a source identical to CUDSS14A derived assuming that
CUDSS14A as observed lies at z = 2 (solid lines starting at the solid
symbols at z = 2) and at z = 4 (dashed lines starting at the open
symbols at z = 4, no V curve is plotted in this case since the Lyman 
break should suppress V very rapidly above z $\sim$4).  The horizontal dashed
area in grey represents the approximate magnitude limits of the HDF. The 
diagonal grey
lines represent more typical limits to ground-based imaging and
are shown for the two sets of magnitude-redshift predictions. 
An $\Omega = 1$ Universe has been assumed.}
 \end{figure}

The IJK colours of CUDSS14A are consistent with an Sb galaxy at z=2.5 
(as suggested in paper II), however the VIJ colours are also consistent 
with an Sa galaxy at z=1.5, and the overall spectrum is so extremely 
red it is not very different from a power law. Therefore we cannot make 
any firm conclusions based on the colours, but most probably z is less 
than 4.5 based on the V detection.

If CUDSS14A were at z $\sim$2.2-2.5 then H$\alpha$ would appear in the range
2.1 to 2.3 $\mu$m,
and the [OIII] 5007 line would appear at 1.60 to 1.75 $\mu$m.  No lines are 
observed. However we note that spectroscopic observations of nearby 
ULIRGs by Kim,
Veilleux and Sanders (1998) imply the lines are suppressed compared to 
the level expected from the FIR-derived star-formation rates in these 
objects, and in fact if the ratio of optical/infrared line emission to 
FIR luminosity in SCUBA-selected galaxies is similar to that found in 
nearby ULIRGs then 8-10 m telescopes will be required to detect them. 
Therefore the non-detection of any lines cannot be used to rule out a 
redshift of 2-3.

The optical/infrared, radio and mm/submm data therefore all argue strongly 
for a redshift 2.0 $< z < 4.5$ with some preference for a redshift around 2.2
to 2.5.

\subsection{AGN Heating ?}

A source at redshift z$>$2 with a dust temperature around 50K has a
minimum possible source size of 0.1 arcsec to produce an observed flux
of 2.9 mJy at 240 GHz, which  at z$\sim$2-5  
corresponds to roughly 1$-$2 kiloparsec, which is the
typical size for circumnuclear starburst regions in local ULIRGs.
Another way of looking at this is to say that if the emitting region in
CUDSS14A {\it is} a typical circumnuclear star-forming region and
it is at z$\sim$2-5, then the submm emission must be quite close to being
optically thick.

Applying the same argument to the question of whether the FIR luminosity 
could be produced by AGN heating, we can say immediately that the 
submm-emitting dust cannot predominantly be heated {\em
directly} by a central AGN, since in order to be at a temperature of 50K or
below, for standard dust properties (e.g. Hildebrand 1983) the grains
would have to be $\sim$ 5 kpc from a central AGN of luminosity 
4$\times$10$^{12}$ L$_{\odot}$ and would therefore have been resolvable.
Conversely if optically thin dust were only 1 kpc from such a central source 
its temperature would
be $\sim$ 200K which is not consistent with the ISO and 450 
$\mu$m  non-detections. However we cannot rule out heating via radiative
 transfer through a large
column of dust, with the observed mm/submm emission arising only close to
the outer surface. 

The launch of the Chandra and XMM X-ray satellites
offer the opportunity to observe AGN emission directly in these sources if
it is present and should help to solve this question definitively. Initial Chandra  observations presented by Mushotsky et al (2000) suggest that the hard x-ray background has been resolved into heavily obscured active galaxies, for which the SCUBA-selected objects are a plausible counterpart. However the limited data so far available from Chandra on SCUBA sources (Fabian et al 2000) suggest that these are more like starburst galaxies, unless they are Compton-thick and the scattered X-rays are weak or also absorbed.

\subsection{Source properties}

We can derive a total
luminosity for CUDSS14A of  $\sim$3-6$\times$10$^{12}$ L$_{\odot}$ for 
2$<$z$<$4.5.  This luminosity estimate is for an
Arp220-like spectrum where the total luminosity is completely dominated
by the Submm/FIR. For an M82-like spectrum the total luminosity could in 
fact be considerably higher. For a standard initial mass function, this
 luminosity corresponds to an
ongoing star-formation rate of $\sim$1000 M$_{\odot}$ yr$^{-1}$, which 
is hard to sustain for a cosmologically significant 
time, unless the IMF is very heavily biased towards massive stars.

We can also estimate a total gas mass using the techniques laid out in
Hildebrand (1983) and Gear (1989) (see also Hughe, Dunlop and Rawlings
1997)  to give roughly $\sim$10$^{10}$ M$_{\odot}$. The uncertainty in this
estimate is probably at least a factor 5 in each direction however, given the
uncertainty in dust temperature, optical depth, grain properties, 
redshift and cosmology. A  galaxy of this mass
could only sustain the derived star formation rate for a period of around
10 million years. So these results are all consistent with CUDSS14A  being
a gas and dust-rich galaxy undergoing a short-lived period of extensive 
star-formation.